\input harvmac
\lref\DES{S. Deser, R. Jackiw, and S. Templeton, Ann. Phys. (N.Y.) {\bf 140},
         372 (1982); J. Hong, Y. Kim, and P.Y. Pac, Phys. Rev. Lett.
         {\bf 64}, 2230 (1990); R. Jackiw and E.J. Weinberg, ibid. {\bf 64},
         2234 (1990).}
\lref\JAC {R. Jackiw and S.Y. Pi, Phys. Rev. Lett. {\bf 64} 2969 (1990);
           R. Jackiw and S.Y. Pi, Phys. Rev. D {\bf 42} 3500 (1990).}
\lref\JKV{R. Jackiw, K. Lee and Erick J. Weinberg, Phys. Rev. {\bf D42},
          3488 (1990); C. Lee, K. Lee and H. Min, Phys. Lett. {\bf B252},
          79 (1990). }
\lref\BAR{D.G. Barci and L.E. Oxman, Phys. Rev. D {\bf 52}, 1169 (1995);
          M.E. Carrington and G. Kunstatter, ibid. {\bf 51}, 1903 (1995). }
\lref\LIB{S. Li and R.K. Bhaduri, Phys. Rev. D {\bf 43}, 3573 (1991).}
\lref\SHIN{Y.M. Cho, J.W. Kim, and D.H. Park, Phys. Rev. D {\bf 45}, 3802
          (1992); J. Shin and J.H. Yee, ibid. {\bf 50}, 4223 (1994);
           C. Duval, P.A. Horvathy, and L. Palla, ibid.
          {\bf 52}, 4700 (1995).  }
\lref\AHA{Y. Aharonov and A. Casher, Phys. Rev. {\bf 19} 2461 (1979). }
\lref\ROS{B. Rosenstein, B. Warr, and S.H. Park, Phys. Rev. Lett.
         {\bf 62} 1433 (1989). }
\lref\FRI{R. Friedberg, T.D. Lee and A. Sirlin, Phys. Rev. {\bf D13}, 2739
(1976), Nucl. Phys. {\bf B115}, 1,32 (1976); R. Friedberg, T.D. Lee,
Phys. Rev. {\bf D15}, 1694 (1977); S. Coleman, Nucl. Phys. {\bf B262}, 263
(1985); K. Lee, J.A. Stein-Schabes, R. Watkins and L.M. Windrow, Phys. Rev.
{\bf D39}, 1665 (1989); Ed Copeland, Edward W. Kolb and K. Lee, ibid.
{\bf 38}, 3023 (1988); C. Kim, S. Kim and Y. Kim, ibid. {\bf 47}, 5434 (1993). }
\lref\FRI{R. Friedberg, T.D. Lee and A. Sirlin, Phys. Rev. {\bf D13}, 2739
(1976), Nucl. Phys. {\bf B115}, 1,32 (1976); R. Friedberg, T.D. Lee,
Phys. Rev. {\bf D15}, 1694 (1977); S. Coleman, Nucl. Phys. {\bf B262}, 263
(1985); K. Lee, J.A. Stein-Schabes, R. Watkins and L.M. Windrow, Phys. Rev.
{\bf D39}, 1665 (1989); Ed Copeland, Edward W. Kolb and K. Lee, ibid.
{\bf D38}, 3023 (1988); C. Kim, S. Kim and Y. Kim, ibid. {\bf D47},
5434 (1993) }
%
              
   \def\e{\epsilon}        
\def\g{\gamma}      \def\k{\kappa}     
  \def\m{\mu}      \def\n{\nu}        
\def\o{\omega}      \def\p{\psi}       
\def\s{\sigma}       
       
%

\def\CH{{\cal H}}   \def\CL{{\cal L}}

%
\def\rd{\partial}

\def\darr#1{\raise1.5ex\hbox{$\leftrightarrow$}\mkern-16.5mu #1}

\def\fr#1#2{{\textstyle{#1\over#2}}}
\def\Fr#1#2{{#1\over#2}}

%

%
\def\title#1#2#3{\Title{#1}{#2}\vskip -0.3in
\centerline{{\titlefont#3}}\vskip 0.3in}

\def\horsea{\centerline{Seungjoon Hyun, Junsoo Shin and Jae Hyung Yee}
\bigskip\centerline{{\it Department of Physics and Institute for Mathematical
Sciences}}
\centerline{{\it Yonsei University}}
\centerline{{\it Seoul 120-749, Korea}}}

\def\horseb{\centerline{Hyuk-jae Lee}
\bigskip\centerline{{\it Department of Physics}}
\centerline{{\it Boston University}}
\centerline{{\it Boston, Massachusetts 02215, U.S.A.}}\vskip 0.3in}
\def\abs{\centerline{{\bf Abstract}}\vskip 0.2in}


%

\Title{\vbox{\baselineskip12pt\hbox{YUMS 96-12}\hbox{SNUTP 96-065}}}
{\vbox{\centerline{Vortex Solutions of Maxwell-Chern-Simons field}
    \vskip2pt\centerline{coupled to 4-Fermion theory }}}

\vskip 0.3in

\horsea
\vskip 0.3in
\horseb

\vskip 0.5in

\abs
We find the static vortex solutions of the model of
Maxwell-Chern-Simons gauge field coupled to a (2+1)-dimensional
four-fermion theory.
Especially, we introduce two matter currents coupled to the
gauge field minimally: the electromagnetic current and
a topological current associated with the electromagnetic current.
Unlike other Chern-Simons solitons the N-soliton solution of this
theory has binding energy and the stability of the solutions
is maintained by the charge conservation laws.

\Date{}

\baselineskip=20pt plus 2pt minus 2pt

Various field theories which include Chern-Simons terms in (2+1)-dimensions
are found to admit interesting classical soliton
solutions \DES \JKV \JAC \LIB \SHIN\ .
Recently it has been found that fermionic field theories coupled to
the Chern-Simons gauge field also admit vortex solutions \LIB\ \SHIN  .
In these models the fermionic fields are coupled to Chern-Simons gauge
field without Maxwell term.
In this note we present a fermion field theory coupled to
Maxwell-Chern-Simons field which possesses interesting vortex solutions
with binding energies.

We consider a four-fermion model coupled to Maxwell-Chern-Simons theory
that interacts electromagnetically in (2+1) dimensions as in Ref.\BAR .
The four-fermion models in (2+1) dimensions are non-renormalizable in the weak
coupling expansion, but is known to be renormalizable
in $\fr{1}{N}$ expansion \ROS , $N$ being the number of flavours.

We consider the model described by the Lagrangian
\eqn\ol{
\CL = -\fr{1}{4} F^{\m \n} F_{\m \n} + \fr{\k}{4} \e^{\m \n \rho} F_{\m \n}
      A_{\rho} + i{\bar \p}^a \g^\m \rd_\m \p^a
      - m {\bar \p}^a \p^a
      + e A_\m ( J^\m + lG^\m ) + \fr{1}{2} g^2 ({\bar \p}^a \p^a )^2 ,
}
where
$$J^\m = {\bar \p}^a \g^\m \p^a,\quad G^\m = \e^{\m \n \rho} \rd_\n J_\rho,$$
index $a$ denotes the fermion flavour running from 1 to N
and $\k$, $l$ and $g$ are coupling constants.
Here, as in Ref.\BAR , we introduce the topological current
$G_\m$, associated with the electromagnetic current $J_{\rho}$,
which describes the induced charge and current density.
We choose $\g$-matrices to be
\eqn\ga{
\g^0 = \s^3, \quad \g^1 = i\s^1, \quad \g^2 = i\s^2.
}

The equations of motion are
\eqn\eoma{
\rd_\n F^{\n \m} + \fr{\k}{2} \e^{\m \n \rho} F_{\n \rho}
= - e ( J^\m + l G^\m ),
}
\eqn\eomb{
\g^\m (i \rd_\m + e A_\m ) \p^a - m \p^a + g^2 \sum_b ( {\bar \p}^b \p^b )
\p^a - el\e^{\m \n \rho} ( \rd_\n A_\m ) \g_\rho \p^a = 0.
}
We choose the gauge $A_0 = 0$ and consider the gauge field
$A_i$ to be static.
By taking the fermion field $\p$
in component form, $\p^a = {\p^a_+ \choose \p^a_- }e^{-iE_f t} $,
the equations of motion
\eomb\ can be written as the coupled equtions for
$\p^a_+$ and $\p^a_-$ ;
\eqn\eomc{\eqalign{
\bigl[ E_f - m + g^2 \sum_b (\mid \p^b_+ \mid^2 - \mid \p^b_- \mid^2 )
- el \e^{ij0} \rd_j A_i \bigr] \p^a_+ &= D_- \p^a_-
- it \bigl[ (\rd_1 - i \rd_2) E_f \bigr] \p^a_- , \cr
\bigl[ -E_f -m + g^2 \sum_b (\mid \p^b_+ \mid^2 - \mid \p^b_- \mid^2 )
+ el \e^{ij0} \rd_j A_i \bigr] \p^a_- &= D_+ \p^a_+
- it \bigl[ (\rd_1 + i \rd_2) E_f \bigr] \p^a_+ ,\cr
}}
where $D_\pm = D_1 \pm i D_2$,  $D_i = \rd_i - ieA_i$.
If we let
\eqn\anza{
\p^a = \p^a_+ {1 \choose 0}e^{-iE_f t}  ,
}
the equations of motion \eomc\ reduce to
\eqn\eome{\eqalign{
(E_f - m + g^2 \rho_+ - el \e^{ij0} \rd_j A_i ) \p^a_+ &= 0, \cr
D_+ \p^a_+ -it \bigl[ (\rd_1 - i\rd_2)E_f \bigr] \p^a_+ &= 0 ,\cr
}}
and if we take
\eqn\anzb{
\p^a = \p^a_- {0 \choose 1}e^{-iE_f t}  ,
}
Eq.\eomc\ becomes
\eqn\eomf{\eqalign{
(E_f + m + g^2 \rho_- - el \e^{ij0} \rd_j A_i ) \p^a_- &= 0, \cr
D_- \p^a_- -it \bigl[ (\rd_1 + i\rd_2)E_f \bigr] \p^a_- &= 0 , \cr
}}
where $J^0_{\pm} = \rho_\pm = \sum_a \mid \p^a_\pm \mid^2$.
The Eqs. \eome\ and \eomf\ show that the fermion fields,
$\p_+$ and $\p_-$, satisfy the
self-duality conditions.

From the form of the solutions \anza\ or \anzb , we find
\eqn\curi{
J^i = {\bar \p}^a \g^i \p^a = 0,
}
which implies that the induced charge also
vanishes;
\eqn\ich{
G^0 = \e^{0ij} \rd_i J_j = 0 .
}
The magnetic fields, for each case of \anza\ and \anzb\ , then reduce to
\eqn\maga{
B_+ = -F_{12} = \Fr{e}{\k} \rho_+ \qquad or \qquad
B_- = -F_{12} = \Fr{e}{\k} \rho_-
}
respectively.

Written in (2+1)-dimentional notation the topological current takes the form
\eqn\icu{
G^i =  \e^{ij0} \rd_j J_0 ,
}
which is related to the induced current \BAR\ by,
\eqn\cur{
G^i_{ind} = l \e^{ij0} \rd_j J_0 .
}
As discussed in Ref.\BAR , this induced current comes
from the magnetic dipole moment density,
\eqn\dip{
\overrightarrow{m} = \fr{\m}{Q} J_0 \hat{z} = l J_0 \hat{z} ,
}
through the relation
$\overrightarrow{G}_{ind} = \overrightarrow{\nabla} \times
\overrightarrow{m}$.
Here, the magnetic dipole moment is $\m \hat{z}$ ($\hat{z}$ is a unit vector
perpendicular to the plane in consideration), $Q$ is a total charge, and
$J_0$ is a charge density.
We note that the field equation \eoma\ can then be written as
\eqn\eoml{
\rd_j F^{ji} = - el \e^{ij0} \rd_j J_0 .
}
By using Eqs. \maga\ and \eoml , we find the constant $l$ to be
$l = -\fr{1}{\k}$.
We thus find that there exists the magnetic dipole moment,
$\m =-\fr{Q}{\k}$, in the system.

By using Eqs. \eome\ and \eomf , the Hamiltonian density $\CH$ reduces to
\eqn\ham{
\CH = \fr{1}{2} F^2_{12} \pm m \rho_{\pm} - \fr{1}{2} g^2 \rho^2_{\pm}
      + \fr{e^2 l}{\k} \rho^2_{\pm} ,
}
where the $\pm$ signs are for the solutions of the forms \anza\ and
\anzb , respectively.
The total energy of the system can be written as
\eqn\ene{\eqalign{
E = \int d^2 r \CH &= \int d^2 r \bigl[ \fr{1}{2} F^2_{12} \pm m \rho_{\pm}
- \fr{1}{2} g^2 \rho^2_{\pm} + \fr{e^2 l}{\k} \rho^2_{\pm} \bigr] \cr
                   &= \int d^2 r \bigl[\pm m \rho_{\pm} + \fr{1}{2}
                     ( \fr{e^2}{\k^2} + \fr{2 e^2 l}{\k} - g^2 ) \rho^2_{\pm}
                     \bigr] . \cr
                   &= \int d^2 r \bigl[\pm m \rho_{\pm} - \fr{1}{2}
                   (\fr{e^2}{\k^2} + g^2 ) \rho^2_{\pm} \bigl] . \cr
}}

The general solutions for the self-duality equation \eome\ and \eomf\ are
well-known\AHA .
To solve the self-dual equations, we note that when $\p_{\pm}$ are
decomposed into its phase and amplitude,
\eqn\deco{
\p_{\pm} = \sqrt{\rho_{\pm}} e^{i\o_{\pm}},
}
where $\p_{\pm} = \sum_a \p^a_{\pm}$,
the self-dual equations \eome\ and \eomf\ can be written as
\eqn\lio{
e \nabla \times \vec A = \pm \nabla \times (\nabla \o_{\pm} ) \mp \nabla
\times (\nabla E_f ) t \mp \fr{1}{2} \nabla^2 \ln \rho_{\pm} .
}
From Eqs. \maga\ and \lio\ we obtain the equation for the charge
density $\rho_{\pm}$:
\eqn\liou{
\nabla^2 \ln \rho_{\pm} = \mp \fr{2e^2}{\k} \rho_{\pm} .
}
Eq.\liou\ is the Liouville equation which is completely integrable.
When we take the solution \anza , $\k > 0$ is required
in order to have nonsingular positive charge density $\rho_+$.
If we take the solution \anzb , however, $\k < 0$ is required for
the nonsingular charge density $\rho_-$.
That is, both solutions involve only one of the (2+1) dimensional spinor
field components, depending on the sign of $\k$.
Therefore, the most general circularly symmetric nonsingular solutions to the
Liouville equations involve two positive constants
$r_\pm$ and ${\cal N}_\pm$ \JAC :
\eqn\solu{
\rho_\pm = \pm \Fr{4 \k {\cal N}^2_\pm}{e^2 r^2} \bigl[ \bigl( \fr{r_\pm}{r}
           \bigr) ^{{\cal N}_\pm} + \bigl( \fr{r}{r_\pm}
           \bigr) ^{{\cal N}_\pm} \bigr] ^{-2} .
}
where $r_{\pm}$ are scale parameters and the ($+$) sign is for the positive
$\k$ and the ($-$) sign for the negative $\k$.
To fix ${\cal N}_\pm$, we observe that regularity at the origin,
$\rho _\pm \ {\buildrel {r \to 0} \over \longrightarrow } \
r^{2{\cal N}_\pm -2}$, and at infinity,
$\rho _\pm \ {\buildrel {r \to \infty } \over \longrightarrow }
 \ r^{-2{\cal N}_\pm -2}$, require ${\cal N}_\pm \geq 1$.
Especially, for single-valued $\p_{\pm}$, ${\cal N}_\pm$ must
be a positive integer.
The charge density $\rho_{\pm}$ are given by the time-component of
$J^\m$ (here, $G^0 = 0$ ). By using Eq.\solu\ the total charge is given by
\eqn\char{
Q_{\pm} = \int \rho_{\pm} d^2 r = \pm \Fr{2\k {\cal N}_{\pm}}{e^2} > 0 ,
}
which implies that the parameter ${\cal N}_\pm$ describes the
total charge of the system.

From the solution \solu , the total energy of the system is shown to be
\eqn\enea{
E = \pm m \Fr{2\k {\cal N_{\pm}}}{e^2} - \bigl[ \bigl( \Fr{e^2}{\k^2}
    + g^2 \bigr) \bigl( \Fr{4\pi \k^2 {\cal N}^3_{\pm} \Gamma(2 -
    \fr{1}{{\cal N_{\pm}}}) \Gamma(2 + \fr{1}{{\cal N_{\pm}}})}
    {3e^4 r^2_{\pm}} \bigr) \bigr] .
}
The total energy \enea\ satisfies the relation
\eqn\energ{
E( {\cal N}_{1} ) + E( {\cal N}_{2} ) + \cdots +
E( {\cal N}_{i} ) \geq E( {\cal N} )
}
where ${\cal N} = {\cal N}_{1} + {\cal N}_{2} + \cdots
+ {\cal N}_{i}$ .
Unlike other Chern-Simons solitons, therefore,
${\cal N}$-soliton solutions of this
model \ol\ have binding energies. In other words, it is energetically
more stable to become ${\cal N}$-soliton than to be in a state of
${\cal N}$ separate single (${\cal N}=1$) solitons.

Because the vortex solutions in this theory are nontopological,
their stability is not guaranteed automatically \JKV\ \FRI .
By using Eq.\curi\ and \ich , however, we see that the electromagnetic
current $J^\m$ and the topological current $G^\m$ are conserved;
\eqn\curr{
\rd_\m J^\m = 0 \qquad and \qquad \rd_\m G^\m = 0 .
}
This implies that $Q_\pm$ and ${\cal N}_\pm$ are conserved; i.e.
\eqn\num{
\dot Q_\pm = \dot {\cal N}_{\pm} = 0 .
}
Thus the stability of the ${\cal N}_\pm$-soliton solutions is guaranteed
by the charge conservation laws of the system.

We note that, in the limit $l \rightarrow 0, \ g \rightarrow 0, \
e^2 \rightarrow \infty$ and $\fr{\k}{e^2} \rightarrow 1$,
the Lagrangian \ol\ reduces to that of Li and Bhaduri \LIB .
It is easy to show that the solutions and the tatal energy \enea\ also
reduce to that of Li and Bhaduri in this limit.
We finally note that the four fermion interaction term does not
affect the structure of the soliton solutions.
In other words, the soliton solutions exist even in the limit
$g \rightarrow 0$, for which case the total energy is given by
\enea\ with $g = 0$.

\vskip 0.3in

\bigbreak\bigskip\bigskip\bigskip\centerline{{\bf Acknowledgements}}
This work was supported in part by the Korea Science and Engineering
Foundation through 95-0701-04-01-3 and 065-0200-001-2,
Center for Theoretical Physics(S.N.U.) and the Basic Science Research
Institute Program, Ministry of Education Project No. BSRI-95-2425.

\listrefs

\end
\bye